\documentclass[12pt]{iopart}

\usepackage{amsfonts}
\usepackage{graphicx}
\errorstopmode

\newcommand{\dd}{\,{\rm d}}
\newcommand{\ii}{{\rm i}}

\newcommand{\ee}{\mbox e}
\newcommand{\ket}[1]{\vert #1\rangle}
\newcommand{\bra}[1]{\langle#1\vert}
\newcommand{\displ}{d}

\begin{document}  

\title{Highly non-Gaussian states created via cross-Kerr nonlinearity}

\author{Tom\'a\v{s} Tyc$^{1,2}$ and Natalia Korolkova$^1$}
\address{${}^1$ School of Physics and Astronomy,
University of St Andrews, North Haugh, St~Andrews, KY16~9SS, Scotland\\
${}^2$Institute of Theoretical Physics, Masaryk
      University, Kotl\'a\v rsk\'a 2, 61137 Brno, Czech Republic}
\ead{tomtyc@physics.muni.cz}

\begin{abstract}
 We propose a feasible scheme for generation of strongly
non-Gaussian states using the cross-Kerr nonlinearity. The resultant states are highly non-classical states of electromagnetic field and exhibit negativity of their Wigner function, sub-Poissonian photon statistics, and amplitude squeezing. Furthermore, the Wigner function has a distinctly pronounced ``banana'' or ``crescent'' shape specific for the Kerr-type interactions, which so far was not demonstrated experimentally. We show that creating and detecting such states should be possible with the present
technology using electromagnetically induced transparency in a four-level atomic system in N-configuration.
\end{abstract}
\pacs{42.65.-k, 42.50.-p, 03.67.-a}

\section{Introduction}

Non-Gaussian and in general non-classical states of light enjoy an increasing
interest in the quantum information community. They are indispensable building
blocks for entanglement generation, distillation and broadcasting, for quantum
information relays and networks in the realm of quantum information processing
based on infinite-dimensional optical systems and atomic ensembles
\cite{Cerf-book,Ill06}. As has been shown, it is impossible to perform optical
quantum computation with linear optics only and a non-linear interaction is
needed to perform a universal quantum computation \cite{Llo99}. Hence it is
unavoidable to investigate the nonlinearities of atomic systems or alternative
solutions for giant nonlinearities and efficient nonlinear coupling, such as
measurement-induced nonlinearities or other non-Gaussian operations.  In
addition, the experimental difficulty associated with the ``on-line'' nonlinear
operations can be circumvent by using the off-line resources, the highly
non-Gaussian states prepared separately from the actual computation and fed
into the computation circuits when necessary to replace the nonlinear gates
\cite{Gho07,Ral03}. Continuous-variable quantum repeaters, quantum relays and a
number of other computation and communication tasks rely on non-Gaussian
operations or non-Gaussian states, states described by a non-Gaussian Wigner
function.

One of the examples of a non-linear coupling is the cross-Kerr effect which
involves controlling the refractive index experienced by one mode of the
electromagnetic field by the intensity of another. In this paper, we propose an
experiment that employs the cross-Kerr effect to create highly non-classical
non-Gaussian states of light via interaction of two coherent beams in an atomic
medium exhibiting electromagnetically-induced transparency \cite{Sch96},
subsequent measurement on one beam and feed-forward on the other. Such
experiment on the one hand responses to the need for an efficient source of
non-Gaussian states, on the other hand can be seen as a test-bench for the
strong non-linear coupling using an EIT based system. In our case, the first
evidence of the large non-linear phase shift in combination with enough
coherence to generate and preserve quantum features can be obtained using
merely a direct photodetection to verify the photon-number squeezing in one of
the output modes. This would pave the way for further applications of such
non-linear systems: Cross-Kerr nonlinear interaction provides a basis for
several proposals of quantum information protocols or their elements, such as
non-demolition photon number detection~\cite{Bea05}, C-NOT gate~\cite{Nem07},
or continuous-variable entanglement concentration
\cite{Fiu03,Men06}.

The first basic ingredient of our scheme is the nonlinear coupling based on the
third-order nonlinearity, an optical cross-Kerr effect \cite{She84}. Such
Kerr-type interaction of the two initially independent coherent beams $a$ and
$b$ entangles them producing a continuous-variable state of modes $a$ and $b$
with quantum correlations between photon number in one mode and phase of the
other. A subsequent measurement done on the mode $b$ generates a conditional
squeezed state of the mode $a$. This is a photon-number squeezed state
described for the first time in \cite{Kit86}. The mechanism behind it is a
re-shaping of the quantum uncertainty due to the Kerr-effect and the influence
of a local measurement on the system of two spatially-separated entangled
modes. A particular feature of the states squeezed using Kerr nonlinearity is
their non-Gaussian character in contrast to the squeezed states produced, e.g.,
in the parametric process based on the nonlinearity of the second order
\cite{Ill06}. The corresponding Wigner function has a specific crescent shape
and exhibits negativity in some regions of the phase space in the form of the
decaying fringes, resembling a section of the Wigner function of the Fock
states. Indeed, there is a strong connection between the states emerging in
this scheme and the Fock states: the generation of the non-classical state in
our scheme can be understood as a superposition effect between different
photon-number or Fock states. Hence the resultant Wigner function can be viewed
a result of interference between the Fock-state Wigner functions with different
photon numbers $n$. We will provide a more detailed discussion of the mechanism
producing this particular state in Sec. 3 and 4 of the paper.

Another inherent feature of our protocol is the controlled displacement. We are dealing with continuous variable quantum systems. It means, that although in each single run of the proposed experiment (corresponding to a single measurement result on mode $b$) a squeezed state is created, the detection of squeezing will not be possible because it requires a statistical processing of an ensemble of measurement results. In a number of subsequent runs, the overall squeezing vanishes, as we then will obtain a mixture of quantum uncertainties with different mean photon numbers and the overall state will become  fuzzy. The way out is provided by the quantum feed-forward: the state in the mode $a$  is displaced in each run with a displacement amplitude determined by a corresponding measurement outcome in mode $b$. This procedure merges all differently displaced crescent-shaped uncertainties into a single crescent which exhibits amplitude (or photon-number) squeezing.

The non-classical non-Gaussian states of this type have not yet been
demonstrated experimentally although the Kerr effect, e.g. in optical nonlinear
fibres, was exploited extensively for squeezing and entanglement generation
(see e.g. \cite{Sil01,Hee03}). The challenging point is to produce the strong
enough nonlinear coupling to generate sensible non-Gaussian features. All the
experiments performed so far were working in the regime of weak nonlinearity
restricting themselves to the first stages of quantum state evolution, which
can be well approximated by the Gaussian Wigner function of the same type as
the one describing squeezing in parametric second-order nonlinear
processes. The most promising candidate for observing the large cross-Kerr
effect is the four-level atomic medium exhibiting electromagnetically induced
transparency (EIT)~\cite{Sch96}. Under certain conditions, the cross-phase
modulation in such media can reach values by many orders of magnitude larger
than it is possible in optical fibres and the phase shift between two photons
can reach values of $10^{-3}$ for hot atoms in a vapour cell or even $10^{-1}$
for cold atoms in a magneto-optical trap~\cite{Wan06} . The experimental
feasibility of achieving a large cross-Kerr interaction for continuous-wave
fields has been demonstrated using cold atoms in magneto-optical
trap~\cite{Kan03}.  Recently, a suitable method was suggested for
group-velocity matched pulses and room temperature atomic gas
cell~\cite{Wan06}. The form of the cross-Kerr interaction in a such four-level
atomic system in the N-configuration can be derived in a simple
way~\cite{Sin07}. Essentially, the strong cross-Kerr interaction is caused by
the ac-Stark shift produced by the third perturbing field to the dark-state of
the lambda subsystem.

The paper is organised as follows. In Sec.~\ref{scheme} we present the
experimental scheme.  Sec.~\ref{state-psi} discusses the non-Gaussian
properties of the state obtained in a single run of the experiment. In
Sec.~\ref{displacement} we show how a controlled displacement on this output
state can reproduce the same highly non-Gaussian state in multiple runs generating a constant output state with high accuracy.  In Sec.~\ref{scaling} we discuss the scaling of the non-Gaussian effects with
experimental parameters and conclude in Sec.~\ref{conclusion}.

%%%%%%%%%%%%%%%%%%%%%%%%%%%%%%%%%%%%%%%%%%%%%%%%%%%%%%%%%%%%%%%%%%%%%%%%
\section{Experimental scheme}
\label{scheme}

\begin{figure}
\begin{center}
\includegraphics[width=80mm]{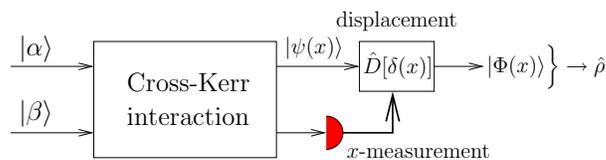}
\caption{Proposed experimental setup for generating crescent states with
  negative Wigner function: two coherent states interact in a cross-Kerr
  medium, one of them is subject to measurement of the field quadrature, and
  displacement is performed on the other one depending on the measurement
  outcome.}
\label{setup}
\end{center}  
\end{figure}

Consider the following experimental scheme sketched in Figure~\ref{setup}: two
coherent states with amplitudes $\alpha$ and $\beta$ occupying modes $a$ and
$b$, respectively, interact in a cross-Kerr medium and mode $b$ is then subject
to a measurement of the $\hat x$ quadrature via homodyne detection.  Based on the
measurement outcome $x$, a displacement operation is performed on mode $a$. In this section we obtain the form of the output state $\rho$ and its properties will be discussed in the following sections.

The Hamiltonian of the cross-Kerr interaction is
\begin{equation}
  \hat H=-\gamma'\hat n_a\hat n_b,
\end{equation}
where $\gamma'$ expresses the strength of the interaction and $\hat n_a=\hat
a^\dagger\hat a$, $\hat n_b=\hat b^\dagger\hat b$ are the photon number
operators of the two modes.  We introduce the phase shift between two photons
as $\gamma=\gamma't$, where $t$ is the interaction time, which gives the
evolution operator $\hat U=\exp(\ii\gamma\hat n_a\hat n_b)$. If the two modes
are originally in coherent states $\ket\alpha_a$ and $\ket\beta_b$, then after
the cross-Kerr interaction their state will be
\begin{eqnarray}\nonumber
  \ket{\Psi}_{ab} = \exp(\gamma\hat n_a\hat n_b)\ket\alpha_a\ket\beta_b
      &=\ee^{-|\alpha|^2/2}\sum_{n=0}^\infty \frac{\alpha^n}{\sqrt{n!}}
    \exp(\gamma\hat n_a\hat n_b)\ket n_a\ket\beta_b \\ 
          &=\ee^{-|\alpha|^2/2}\sum_{n=0}^\infty \frac{\alpha^n}{\sqrt{n!}}
      \ket n_a \,\ket{\beta\ee^{\ii\gamma n}}_b
\label{superposition}\end{eqnarray}
This is an entangled state where the photon number $n_a$ in mode $a$ is
correlated with the amplitude of coherent state in mode $b$ by introducing to it a phase shift dependent on $n_a$ (and vice versa).

In the next step the quadrature $\hat x=(\hat b+\hat b^\dagger)/\sqrt2$ of mode
$b$ is measured via homodyne detection with the outcome $x$, which results in
the following (unnormalized) state of mode $a$:
\begin{equation}
  \hspace*{-22mm} \ket{\psi(x)}_a={}_b\bra{x} \Psi\rangle_{ab} 
  =\frac{\ee^{-|\alpha|^2/2}}{\sqrt[4]\pi}
   \sum_{n=0}^\infty \frac{\alpha^n
   \exp[-(x-\sqrt2\beta_n)^2/2+\ii\sqrt2\beta'_nx-\ii\beta_n\beta'_n]}
   {\sqrt{n!}} \,\, \ket n_a
\label{entangled}
\end{equation}
where we have denoted $\beta_n={\rm Re}\,(\beta\ee^{\ii\gamma n})$,
$\beta'_n={\rm Im}\,(\beta\ee^{\ii\gamma n})$ and used the expression for a
coherent state in $x$-representation.  The square of the norm of
$\ket{\psi(x)}_a$ is the probability density $P(x)$ that the particular
measurement outcome $x$ occurs.

As the last step, a displacement in the phase space is applied to mode $a$. The
resulting state of mode $a$ is then
\begin{equation}
 \ket{\Phi(x)}=\hat D[\displ(x)]\ket{\psi(x)}=\ee^{\displ(x)\hat
 a^\dagger-\displ^*(x)\hat a}\ket{\psi(x)},
\label{Phi}\end{equation}
where $\displ(x)$ is the displacement parameter that depends on the measurement
outcome $x$.
When the experiment is performed repeatedly, then the output state is averaged
to
\begin{equation}\label{rho}
  \hat\rho=\int_{\mathbb R} \ket{\Phi(x)} \bra{\Phi(x)}\dd x,
\end{equation}
which is properly normalized.

The state $\ket{\Phi(x)}$ in general depends on the measurement outcome
$x$. However, we will show that in a certain regime this dependence can be
quite weak.
This way, by running the experiment many times (and obtaining different values
of $x$), one can still get almost a pure state $\hat\rho$ at the output mode
$a$ that has highly non-classical properties.

%%%%%%%%%%%%%%%%%%%%%%%%%%%%%%%%%%%%%%%%%%%%%%%%%%%%%%%%%%%%%%%%%%%%%%%%
\section{Properties of the state $\ket{\psi(x)}$}
\label{state-psi}

Before describing the averaged output state $\hat\rho$, in this section we will
focus on the state $\ket{\psi(x)}$ of mode $a$ after the $\hat x$ measurement
on mode $b$ has been performed. Consider the specific situation of
$\beta=\ii|\beta|\ee^{-\ii\gamma|\alpha|^2}$, i.e., when the phase of of the
mode $b$ is set to $\pi/2-\gamma|\alpha|^2$. We also put a constraint on
$|\alpha|$, such that $|\alpha|\le10$ and the product $|\alpha\beta|\gamma$ is
of order of unity.  The reasons for these assumptions will be explained later
in this section. Taking into account that $\gamma$ is several orders of
magnitude smaller than unity, it also holds that $\gamma|\alpha|\ll1$.

\begin{figure}[h]
\begin{center}
 \includegraphics[width=5cm, angle=270]{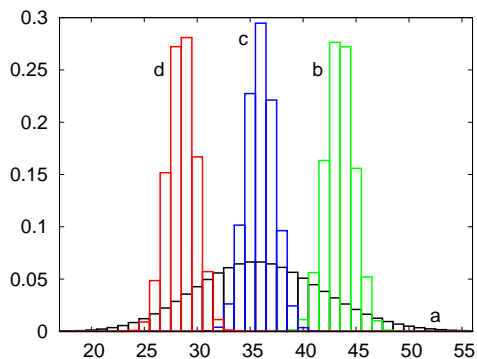}
\caption{Photon number probability distribution of the coherent state
 $\ket\alpha$ (a, black curve) compared with the normalized distributions of
 the states $\ket{\psi(x)}$. The parameters are $|\alpha|=6$,
 $\gamma|\beta|=0.36$ and $x=-4$ (b, green curve), $x=0$ (c, blue curve) and
 $x=4$ (d, red curve).  The photon number distribution in the states
 $\ket{\psi(x)}$ is about $4\times$ squeezed compared to the Poissonian
 statistics of the coherent state $\ket\alpha$.}  \label{fock}
\end{center}
\end{figure}

\begin{figure}[h] 
\begin{center}
\includegraphics[width=65mm]{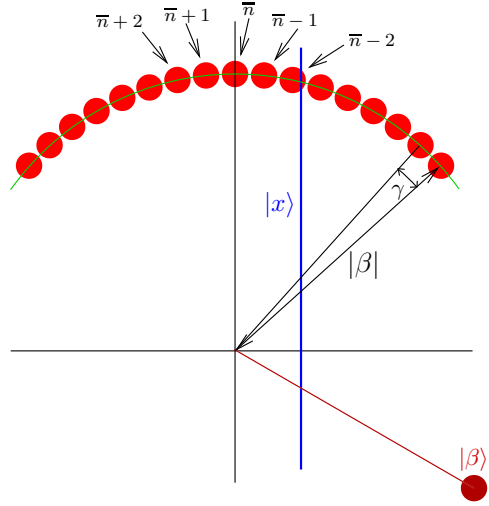}
\caption{Illustration in the phase space of how the photon number squeezing of
the state $\ket{\psi(x)}$ emerges: the red circles represent the uncertainty
regions of coherent states $\ket{\ee^{\ii\gamma n}\beta}_b$ in
(\ref{superposition}) entangled with the photon numbers $n$ in mode $a$
($\overline n\approx|\alpha|^2$ denotes the most likely photon number).  The
circles are distributed along the green arch centred at the origin of the
phase space, having the radius $|\beta|$ and stretching to the angle of
approximately $\gamma|\alpha|$ to both sides of the imaginary axis.  If then
the quadrature $\hat x$ is measured on mode $b$ with a particular outcome $x$
(blue line represents the quadrature eigenstate $\ket x$), then only
those states are chosen from the superposition (\ref{entangled}) that have a
non-negligible overlap with $\ket x$.  If the length of the arch,
$2\gamma|\alpha\beta|$, is larger than the spatial extension of the uncertainty
region of about unity, then only a relatively small number of circles have a
significant overlap with $\ket x$ and so the photon number variance in the
state $\ket{\psi(x)}$ is reduced significantly compared to the state
$\ket\alpha$.
The effect is the strongest when the arch is, very loosely speaking,
perpendicular to the line representing $\ket x$; this gives the phase condition 
on $\beta$.}
\label{oblouk}
\end{center} 
\end{figure}

In this particular case, for evaluating the real and imaginary parts of
$\beta\ee^{\ii\gamma n}$ to calculate $\beta_n$ and $\beta'_n$ in
(\ref{entangled}), we use the expansion of the exponential function as
follows:
\begin{equation}\label{approx}
  \beta\ee^{\ii\gamma n}=\ii|\beta|\ee^{\ii\gamma(n-|\alpha|^2)}
     \approx\ii|\beta|-\gamma|\beta|(n-|\alpha|^2).
\end{equation}
We terminated the series after the second term since $\gamma(n-|\alpha|^2)\ll1$
for all $n$ for which the probability of having $n$ photons in the state
$\ket\alpha$ is non-negligible. Indeed, the photon number distribution of the
state $\ket\alpha$ is Poissonian with both mean and variance equal to
$|\alpha|^2$, hence $(n-|\alpha|^2)$ is of order of $|\alpha|$ and therefore
$\gamma(n-|\alpha|^2)\ll1$ holds in agreement with our assumption
$\gamma|\alpha|\ll1$.

\begin{figure}
\begin{center}
 \includegraphics[width=9cm, angle=0]{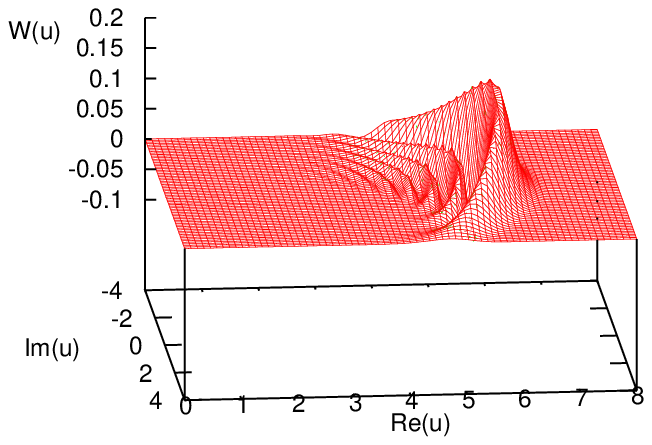}
 \includegraphics[width=9cm, angle=0]{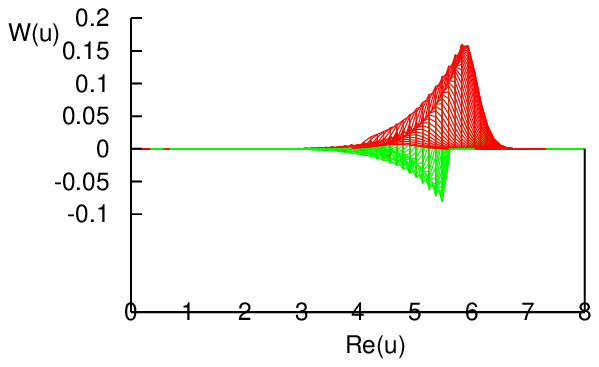}
\caption{Two different views of the Wigner function $W(u)$ of the output state
  $\ket{\psi(x)}$ for $|\alpha|=6$, $\gamma|\beta|=0.36$ and  $x=0$.  The
  product $|\alpha\beta\gamma|$ is equal to 2.16 and the phase of $\alpha$ is
  set to $-\gamma|\beta|^2$ to make the phase of the output state zero. The
  crescent shape and negative values of the Wigner function are clearly
  visible.}
\label{wig1}
\end{center}  
\end{figure}

\begin{figure}
\begin{center}
 \includegraphics[width=9cm, angle=0]{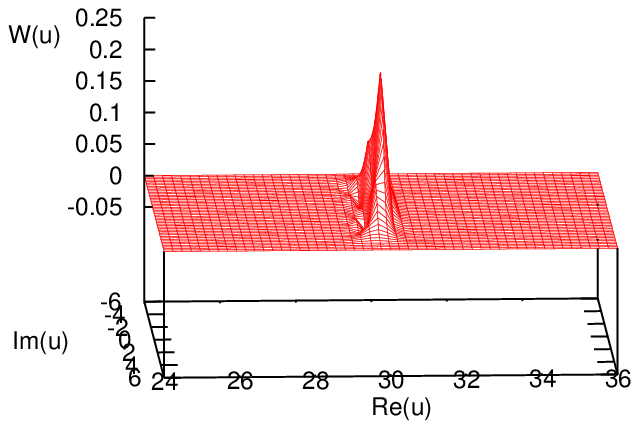}
 \includegraphics[width=9cm, angle=0]{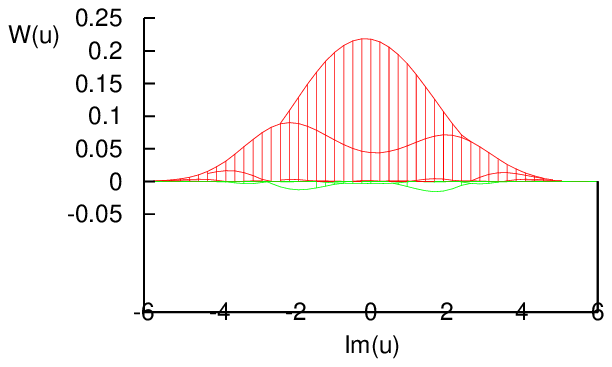}
\caption{Two different views of the Wigner function  $W(u)$ of the output state
$\ket{\psi(x)}$ for $|\alpha|=30$, $\gamma|\beta|=0.066$ and $x=0$ such that
$|\alpha\beta\gamma|= 1.98$. The state is strongly squeezed and is close to a
Gaussian state, hence the negativeness of the Wigner function is weak.}
\label{wig3}
\end{center}  
\end{figure}

From (\ref{approx}) we get $\beta_n=-|\beta|\gamma(n-|\alpha|^2)$ and
$\beta'_n=|\beta|$. Substituting this into  (\ref{entangled}) we arrive at
\begin{equation}  
 \hspace*{-20mm}  \ket{\psi(x)} = 
\frac{\ee^{\ii\phi-|\alpha|^2/2}}{\sqrt[4]\pi}
 \sum_{n=0}^\infty \frac{\alpha^n}{\sqrt{n!}}\,
    \exp\left[\ii \gamma|\beta|^2n-\gamma^2|\beta|^2\left(n-|\alpha|^2
    +\frac x{\sqrt2|\beta|\gamma}\right)^2\right]  \,\ket n,
\label{psix}
\end{equation}
where $\phi=\sqrt2\,x|\beta|+\gamma|\alpha\beta|^2$ is an irrelevant global
phase factor. The important term here is the exponent, which provides the
insight into the main physical effects due the cross-Kerr interaction. The
first term in the exponent shows that the phase of the state $\ket{\psi(x)}$ is
increased by $\gamma|\beta|^2$ with respect to the phase of the original state
$\ket\alpha$.  The second, Gaussian term causes the photon number distribution
in $\ket{\psi(x)}$ to be altered with respect to the original Poissonian
distribution of the state $\ket\alpha$: If $2|\alpha\beta|\gamma>1$, then this
term becomes more important than the factor $\alpha^n/\sqrt{n!}$. Moreover, as
the product $2|\alpha\beta|\gamma$ grows above unity, the photon number
distribution quickly gets dominated by the Gaussian term. In this situation the
mean photon number is $|\alpha|^2-x/(\sqrt2|\beta|\gamma)$ compared to
$|\alpha|^2$ for the coherent state $\ket\alpha$ and the variance of the photon
number distribution is $1/(2\gamma|\beta|)^2$ compared to $|\alpha|^2$ for
$\ket\alpha$.  This means that the state $\ket{\psi(x)}$ is squeezed in the
photon number by the factor of approximately $2|\alpha\beta|\gamma$, as
illustrated in Figure~\ref{fock} for $|\alpha\beta|\gamma\approx 2.16$ and
different values of the measurement result $x$. Thus if
$2|\alpha\beta|\gamma>1$, then the output state will exhibit a strongly
sub-Poissonian photon statistics.  Figure~\ref{oblouk} gives a visualisation of
the above argumentation. Equation~(\ref{psix}) and Figure~\ref{oblouk} also
provide an argument for the particular choice of the phase of the mode $b$,
$\beta=\ii|\beta|\ee^{-\ii\gamma|\alpha|^2}$.  This way we have demonstrated
the photon number variance reduction -- photon number squeezing in the output
state $\ket{\psi(x)}$ obtained in a single run of the experiment.  A
consequence of this reduction are several non-classical phenomena: negativity
and crescent shape of the Wigner function, as well as amplitude squeezing.

Figures~\ref{wig1} and \ref{wig3} show the Wigner function
corresponding to the output state $\ket{\psi(x)}$ of  (\ref{psix}) for
different values of $|\alpha|$ and $\gamma|\beta|$. The above mentioned
features of the Wigner function are clearly visible especially in
Figure~\ref{wig1}. 
Comparison of Figures~\ref{wig1} and \ref{wig3} provides a clear insight into the
assumptions made in the beginning of this section. For large values of
$|\alpha|$ (say over 10) one gets a state that is close to a squeezed Gaussian
state and the features we are looking for, namely the negativity and
crescent-shape of the Wigner function, are suppressed (Figure~\ref{wig3}). We
also assumed that the product $|\alpha\beta|\gamma$ is of order of unity
because otherwise the photon number squeezing is small, the state
$\ket{\psi(x)}$ is close to a coherent state and hence does not have strong
non-Gaussian and non-classical properties. $|\alpha\beta|\gamma>1$ requires a
large cross-Kerr nonlinearity and/or a large amplitude $\beta$. The former we
anticipate to be feasible with the four-level EIT based system mentioned in the
Introduction. The Wigner function in Figure~\ref{wig1} is plotted for the
parameters satisfying all the discussed restrictions, with
$|\alpha\beta|\gamma\approx 2.16$ as in Figure~\ref{fock}. Such features as the
negativity and crescent-shape are highly pronounced, confirming the strongly
non-Gaussian character of the output state.

%%%%%%%%%%%%%%%%%%%%%%%%%%%%%%%%%%%%%%%%%%%%%%%%%%%%%%%%%%%%%%%%%%%%%%%%%%%%%%
\section{Effect of displacement in the phase space}
\label{displacement}

As we have seen, the output state $\ket{\psi(x)}$ depends on the value $x$
obtained by the measurement. Specifically, for different $x$, there will be
different mean photon numbers in the output state of the mode $a$.  To prepare an
identical crescent state repeatedly, one strategy would be to post-select the
output at some sufficiently narrow interval of $x$, which would, however,
reduce the success rate significantly. Another option, however, is to correct
for the change of the mean photon number caused by the $x$-measurement.  It
turns out that this correction can be achieved with a good accuracy by
performing a phase-space displacement operation $\hat D[(\displ(x)]$
with displacement parameter $\displ$ depending on $x$. In this way one can
obtain a state $\ket{\Phi(x)}$ almost independent of the measurement outcome
$x$, which makes $\hat\rho$ in  (\ref{rho}), Figure~\ref{setup} nearly a pure state.

To find the displacement magnitude $|\displ(x)|$ that would compensate for the
change of the mean photon number, we recall that for coherent states, the mean
photon number depends quadratically on the amplitude.  The
state $\ket{\psi(x)}$ has the mean photon number
$|\alpha|^2-x/(\sqrt2|\beta|\gamma)$ and therefore its
mean amplitude $|\alpha'|$ is equal to the square root of this number. In
order to revert to the original amplitude $|\alpha|$, we need to displace by $|\alpha|
-|\alpha'|$.  The direction of the displacement in the phase space is given by
the phase of the state $\ket{\psi(x)}$, which is, as seen from
 (\ref{psix}), equal to $\gamma|\beta|^2+\arg(\alpha)$.  This way we arrive
at the displacement parameter
\begin{equation}
  \displ(x)=\left(|\alpha|-\sqrt{|\alpha|^2-\frac{x}{\sqrt
      2\,\gamma|\beta|}}\right) \,\ee^{\ii[\gamma|\beta|^2+\arg(\alpha)]} .
\label{delta}
\end{equation}

\begin{figure}
\begin{center}
\includegraphics[width=40mm,angle=270]{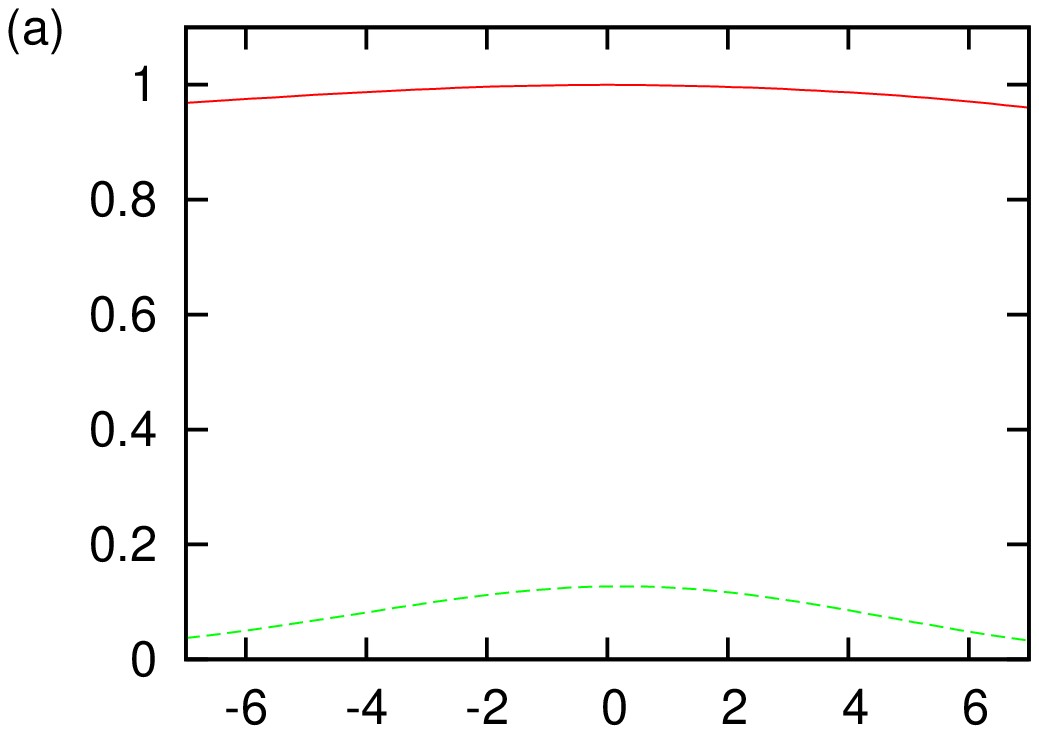}
\includegraphics[width=40mm,angle=270]{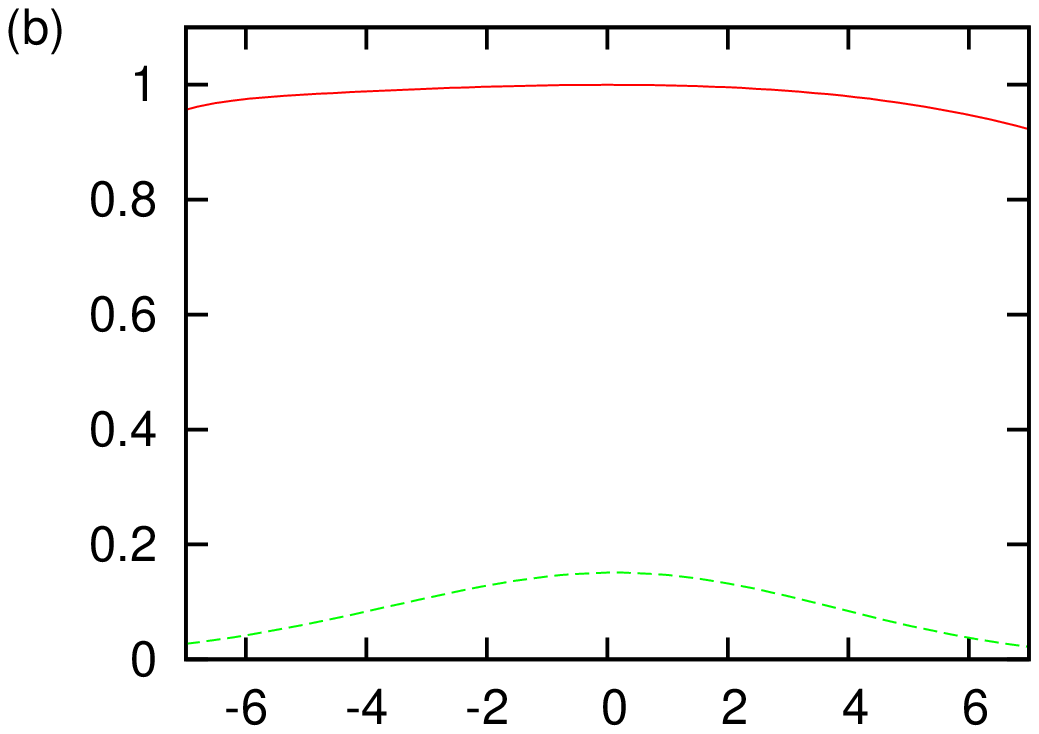}
\caption{The normalized scalar product $F(x)$ (solid red line) and the
probability density $P(x)$ (dashed green line) as a function of $x$ for (a)
 $|\alpha|=6, \gamma|\beta|=0.36$ and (b) $|\alpha|=9,
 \gamma|\beta|=0.2$. Since the function $F(x)$ is much broader than $P(x)$, the
 fidelity $F(x)$ is almost unity for all values $x$ likely to be found in the
 measurement, and therefore the state $\ket{\Phi(x)}$ is almost independent of
 $x$.}
\label{scalar_product}
\end{center}  
\end{figure}

To see that the state $\ket{\Phi(x)}$ in  (\ref{Phi}) is really almost
independent of the measurement outcome $x$, we investigate the normalized
scalar product
\begin{equation}
     F(x)\equiv\frac{\langle\Phi(0)\ket{\Phi(x)}}
         {\sqrt{\langle\Phi(0)\ket{\Phi(0)}\langle\Phi(x)\ket{\Phi(x)}}}
\label{}
\end{equation}
that expresses the overlap of the displaced output state corresponding to an
arbitrary $x$ and the one corresponding to $x=0$.  Figure~\ref{scalar_product}
shows the function $F(x)$ and the probability density
$P(x)=\langle\psi(x)\ket{\psi(x)}$ of the quadrature for a few values of
$|\alpha|$ and $\gamma|\beta|$. It can be seen that in the regions of $x$ where
the probability $P(x)$ is non-negligible, the function $F(x)$ has values close
to unity, and this is a typical behaviour. This means that the state
$\ket{\Phi(x)}$ is close to $\ket{\Phi(0)}$ for all $x$ that are likely to be
found in the quadrature measurement. Therefore $\hat\rho$, which is the mixture
of these states [see  (\ref{rho})], will be almost a pure state. Numerical
simulations confirm this and the purity ${\cal P}={\rm Tr}\,\hat\rho^2$ has
values of approximately 0.95 for the states corresponding to parameters
$\alpha,\beta$ and $\gamma$ in Figures~\ref{wig1} and \ref{wig3}.

\begin{figure}
\begin{center}
\includegraphics[width=65mm]{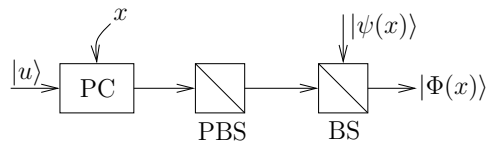}
\caption{A possible way of realizing the displacement operation: a vertically
  polarised coherent beam is directed into a Pockels cell (PC) with the optical
  axis rotated by $45$ degrees and controlled by the outcome $x$ of quadrature
  measurement. The output beam then impinges onto a polarising beam splitter
  (PBS), which yields a polarised coherent beam with amplitude depending on
  $x$. This beam is then mixed with the state $\ket{\psi(x)}$ on a highly
  reflective beam splitter (BS). The displaced state $\ket{\Phi(x)}$ is found
  at one output.  }
\label{displ_exper}
\end{center}  
\end{figure}

Experimentally, the desired controlled displacement can be realized mixing the
state $\ket{\psi(x)}$ with a coherent state $\ket{\displ/t}$ on a beam splitter
with a very low transmissivity $t$. If $|t|\ll1$, then one obtains almost
exactly the state $\hat D(\displ)\ket{\psi(x)}$ at one beam splitter output
port~\cite{Fur98}. The amplitude $\displ/t$ can be varied electronically
according to the measured value $x$ and  (\ref{delta}) using, e.g., the
scheme depicted in Figure~\ref{displ_exper}.

%%%%%%%%%%%%%%%%%%%%%%%%%%%%%%%%%%%%%%%%%%%%%%%%%%%%%%%%%%%%%%%%%%%%%%%%
\section{Scaling of the non-classical effects with $|\alpha|,|\beta|$ and
  $\gamma$}
\label{scaling}

In what follows, we discuss the scaling of the effects described above for the
output state $\ket{\Phi(0)}=\ket{\psi(0)}$, as we have shown that the state
$\ket{\Phi(x)}$ is almost independent of $x$.  An inspection of formulas
(\ref{psix}) and (\ref{delta}) shows that, apart from phase factors, the form
of the output states depends on $|\alpha|$ and the product $\gamma|\beta|$.

Consider first the effect of varying $\gamma|\beta|$ while keeping $|\alpha|$ constant.
For $\gamma|\beta|=0$, the output state reduces to the
original state $\ket\alpha$.  For very small $\gamma|\beta|$ for which
$\gamma|\alpha\beta|\ll 1$, the state $\ket{\psi(0)}$ is close to a weakly
squeezed coherent state with amplitude $\alpha\exp(\ii\gamma|\beta|^2)$, so it
is almost Gaussian. As $\gamma|\beta|$ increases, $\ket{\psi(0)}$ starts to deviate from
a Gaussian state, negative regions of the Wigner functions get larger, and the
crescent shape of $\ket{\psi(0)}$  emerges, as well as the photon number squeezing. When
$\gamma|\beta|$ gets larger than unity, $\ket{\psi(x)}$ is almost a Fock state
$\ket n$ with photon number $n$ depending on the measured value of $x$, and it
is no longer true that $\ket{\Phi(x)}$ is almost independent of $x$.
In this situation our scheme (without the displacement operation) can be used
as a conditional source of Fock states.

The effect of varying $\alpha$ while keeping the product $\gamma|\alpha\beta|$
constant is different. The phase of $\alpha$ influences just the phase of
$\ket{\psi(0)}$.  If $|\alpha|$ is small, the crescent shape of the Wigner
function is strongly pronounced, as well as its negativeness, because the state is squeezed in
photon number and is not far from a Fock state. Close to the origin of the
phase space the circles corresponding to a fixed photon number are more curved
than those farther from the origin, and hence the Wigner function corresponding
to smaller $|\alpha|$ shows a stronger crescent shape than the one
corresponding to larger $|\alpha|$. In contrast, for larger $|\alpha|$ the state is
closer to a Gaussian state whose Wigner function is positive. Hence the
negativeness of the Wigner function is more profound in case of smaller
$|\alpha|$, and this case is more appealing experimentally. Figures ~\ref{wig1} -- \ref{wig3} illustrate this behaviour.

%%%%%%%%%%%%%%%%%%%%%%%%%%%%%%%%%%%%%%%%%%%%%%%%%%%%%%%%%%%%%%%%%%%%%%%%%%%%%%
\section{Conclusion}
\label{conclusion}

In conclusion, we have suggested a feasible scheme to generate highly
non-Gaussian states exhibiting crescent-shaped Wigner function with negative
regions. We envisage the application of such states in quantum information
processing using infinitely-dimensional, continuous-variable quantum
systems. The non-Gaussian operations and non-Gaussian states attract currently
an increasing attention in this context (see e.g. \cite{Gho07,NonGaus} and
references therein) and there is a high demand in the community for their
experimentally viable implementation.  In our scheme, the initial states as
well as the detection (Figure~\ref{setup}), feed-forward
(Figure~\ref{displ_exper}) and evaluation steps belong to the standard toolbox
of quantum optics and are readily available in the laboratory. The first
verification of quantum features can be performed with merely direct
photodetection to prove the photon-number squeezing of Fig~\ref{fock}. The
negativity of the Wigner function and its crescent shape can be visualised
using quantum state tomography \cite{ulf}, a more involved but established
procedure \cite{Lvo07,Bre97}. For the first demonstration of the effect using
the tomographic measurement of the Wigner function of the state $\hat\rho$, the
displacement operation of Figure~\ref{displ_exper} can be simplified via
replacing it by the electronic one.

The challenging part of the proposed scheme is the strong nonlinear coupling
implied. Again, it represents an important building block in quantum
information processing attracting recently a substantial interest from both the
theoretical and experimental sides. A feasible nonlinear coupling device would
make a profound impact on the development of quantum communication and
computation protocols. So far, there are only few first quite involved
implementations demonstrating this effect \cite{Kan03}, \cite{Roo07}. For the
realization of our scheme, we suggest the nonlinear optical cross-Kerr effect
in an EIT based four-level atomic system, which was closely studied recently
both theoretically \cite{Sch96,Wan06,Sin07,Bea04} and experimentally
\cite{Kan03}. Furthermore, due to the experimental availability of the other
elements and particular simplicity of the first verification using direct
photodetection, our suggested scheme can serve as a test bench for the strong
nonlinear coupling preserving quantum effects.

%%%%%%%%%%%%%%%%%%%%%%%%%%%%%%%%%%%%%%%%%%%%%%%%%%%%%%%%%%%%%%%%%%%%%%%%
\section*{Acknowlegments}
We thank Friedrich  K\"onig and Chris Kuklewicz for valuable discussions.  We
gratefully acknowledge the financial support of the EU project COVAQIAL
(FP6-511004) under STREP and of the Leverhulme Trust.\\ \\

\end{document}